\documentstyle[aps,epsf]{revtex}
\begin{document}
%--------------------------------------------------------------------
\def\Paris{Par\'\i{}s}
%--------------------------------------------------------------------
\def\today{\space\number\day\ \ifcase\month\or
  January\or February\or March\or April\or May\or June\or
  July\or August\or September\or October\or November\or December\fi
  \ \number\year}
%--------------------------------------------------------------------
\twocolumn[\hsize\textwidth\columnwidth\hsize\csname@twocolumnfalse\endcsname

\title{Geometrodynamics of Variable-Speed-of-Light Cosmologies}
\author{Bruce A. Bassett,$^{\dagger,*,a}$
        Stefano Liberati,$^{\S,**,b}$
        Carmen Molina--\Paris,$^{\P,c}$
        and Matt Visser$^{\ddagger,d}$}
\address{$^{\dagger}$ Relativity and Cosmology Group, 
         Division of Mathematics and Statistics, \\
         Portsmouth University,~PO1~2EG, England\\
         ${}^*$Department of Theoretical Physics, 
         University of Oxford, 1 Keble Road, OX1 3NP, England\\         
         $^{\S}$International School for Advanced Studies (SISSA),
         Via Beirut 2--4, 34014 Trieste, Italy\\
         $^{**}$Istituto Nazionale di Fisica Nucleare (INFN), 
         sezione di Trieste, Italy\\
         $^{\P}$Theoretical Division, Los Alamos National Laboratory, 
         Los Alamos, New Mexico 87545, USA\\
         $^{\ddagger}$Physics Department, 
         Washington University,
         Saint Louis, Missouri 63130-4899, USA
        }
\date{25 January 2000; Revised 29 June 2000; \LaTeX-ed \today}
\maketitle

\vspace*{0.2cm}
\centerline{Physical Review D, in press}
\vspace*{0.2cm}

\vspace*{0.4cm}
\centerline{\em This paper is dedicated to the memory of Dennis Sciama}
\vspace*{0.4cm}

%--------------------------------------------------------------------

\begin{abstract}
Variable-Speed-of-Light (VSL) cosmologies are currently attracting
interest as an alternative to inflation.  We investigate the
fundamental geometrodynamic aspects of VSL cosmologies and provide
several implementations which do not explicitly break Lorentz
invariance (no ``hard'' breaking).  These ``soft'' implementations of
Lorentz symmetry breaking provide particularly clean answers to the
question ``VSL with respect to what?''.  The class of VSL cosmologies
{\em we consider} are compatible with both classical Einstein gravity
and low-energy particle physics.  These models solve the ``kinematic''
puzzles of cosmology as well as inflation does, but {\em cannot} by
themselves solve the flatness problem, since in their purest form no
violation of the strong energy condition occurs.
We also consider a heterotic model (VSL  plus
inflation) which provides a number of observational
implications for the low-redshift universe if $\chi$ contributes to
the ``dark energy'' either as CDM or quintessence. These implications
include modified gravitational lensing, birefringence, variation of
fundamental constants and rotation of the plane of polarization of
light from distant sources.
\end{abstract}

\bigskip
\hbox to 6.25 in{\hfil PU-RCG-99/25}
\bigskip
]

%--------------------------------------------------------------------
\def\M{{\cal M}}
\def\I{{\cal I}}
\def\L{{\cal L}}
\def\dM{\partial{\cal M}}
\def\implies{\Rightarrow} 
\def\VSL{{$\chi$VSL}}
%--------------------------------------------------------------------
\def\Astrofisica{Astrof\'\i{}sica}
\def\Fisica{F\'\i{}sica}
%--------------------------------------------------------------------
\def\etal{\emph{et al.}}
\def\etc{\emph{etc}}
\def\ie{\emph{i.e.}}
\def\eg{\emph{e.g.}}
\def\aka{\emph{aka}}
%--------------------------------------------------------------------
\def\beq{\begin{equation}}
\def\eeq{\end{equation}}
\def\beqn{\begin{eqnarray}}
\def\eeqn{\end{eqnarray}}
\def\com{{\cal\omega}}
%-----------------------------------------------------------------
\def\cp{c_{\mathrm photon}}
\def\cm{c_{\mathrm matter}}
\def\cn{c_{\mathrm gravity}}
\def\cg{c_{\mathrm gravity}}
%------------------------------------------------------------------
\def\gem{g_{\mathrm em}}
\def\ggrav{g_{\mathrm gravity}}
%------------------------------------------------------------------
\def\gravity{{\mathrm gravity}}
\def\emag{{\mathrm em}}
\def\normal{{\mathrm normal}}
\def\matter{{\mathrm{matter}}}
%-------------------------------------------------------------------
\def\Tr{\hbox{Tr}}
\def\sign{\hbox{sign}}
\def\implies{\Rightarrow}
%--------------------------------------------------------------------
\def\HRULE{\bigskip \hrule \bigskip}
%--------------------------------------------------------------------
\section{Introduction} 
%--------------------------------------------------------------------

High-energy cosmology is flourishing into a subject of observational
riches but theoretical poverty. Inflation stands as the only
well-explored paradigm for solving the puzzles of the early universe.
This monopoly is reason enough to explore alternative scenarios and
new angles of attack.  Variable-Speed-of-Light (VSL) cosmologies have
recently generated considerable interest as alternatives to
cosmological inflation which serve both to sharpen our ideas regarding
falsifiability of the standard inflationary paradigm, and also to
provide a contrasting scenario that is hopefully amenable to
observational test. 

The major variants of VSL cosmology under
consideration are those of Moffat~\cite{Moffat93a,Moffat93b,Moffat98},
Ellis-Mavromatos-Nanopoulos~\cite{EMN}, Clayton and
Moffat~\cite{Clayton-Moffat-1,Clayton-Moffat-2}, and
Albrecht--Barrow--Magueijo~\cite{Albrecht98,Barrow98a,Barrow98b,%
Barrow98c,Albrecht99}, plus more recent contributions by Avelino and
Martins~\cite{Avelino-Martins}, Drummond~\cite{Drummond},
Kiritsis~\cite{Kiritsis}, and Alexander~\cite{Alexander}. The last two
are higher-dimensional, brane-inspired implementations. For
completeness we also mention the earlier work by Levin and
Freese~\cite{LF} which discussed the inflationary-type cosmologies
resulting from a dynamical Planck's constant.

The covariance of General Relativity means that the set of
cosmological models consistent with the existence of the apparently
universal class of preferred rest frames defined by the Cosmic
Microwave Background (CMB) is very small and non-generic.  Inflation
alleviates this problem by making the flat
Friedmann--Lemaitre--Robertson--Walker (FLRW) model an attractor
within the set of almost--FLRW models, at the cost of violating the
strong energy condition (SEC).  Most of the above quoted VSL
cosmologies, by contrast, sacrifice (or at the very least, grossly
modify) Lorentz invariance at high energies, again making the flat
FLRW model an attractor.  In contrast, we will see that the {\em
``soft breaking''} prescription we advocate cannot solve the flatness
problem without additional external sources of energy condition
violation, despite recent claims to the contrary (see
section~\ref{moffat} for details).

In this paper we want to focus on some basic issues in VSL cosmology
that are to our minds still less than clear.  In particular, we wish
to answer the question ``Can we have VSL without explicitly violating
Lorentz invariance?''  As we will see, our approach is to split the
degeneracy between the (effective) null cones of various species of
particles. This means that in {\em our implementations} of VSL
cosmology the Lorentz symmetry is broken in a ``soft'' manner, rather
than in a ``hard'' manner. This ``soft'' breaking of Lorentz
invariance, due to the nature of the ground state or initial
conditions, is qualitatively similar to the notion of spontaneous
symmetry breaking in particle physics, whereas ``hard'' breaking,
implemented by brute force, is qualitatively similar to the notion of
explicit symmetry breaking in particle physics.

We shall have little specific to say about ``hard''
breaking, in the style of Albrecht--Barrow--Magueijo, other than to
point out that ``hard'' breaking is a rather radical modification of
standard physics. In comparison, ``soft'' breaking is rather benign and 
is easier to formulate in a geometrodynamic manner, as we discuss in section
(\ref{2metric}).

We specifically want to assess the geometric consistency of the
VSL idea and ask to what extent it is compatible with Einstein
gravity. This is not a trivial issue: Ordinary Einstein gravity has
the constancy of the speed of light built into it at a fundamental
level; $c$ is the ``conversion constant'' that relates time to space.
We need to use $c$ to relate the zeroth coordinate to time: $d x^0 = c
\; dt$. Thus, simply replacing the \emph{constant} $c$ by a
position-dependent {\emph{variable}} $c(t,\vec x)$, and writing $dx^0
= c(t,\vec x)\; dt$ is a suspect proposition.  Indeed, even the
choice $dx^0 = c(t,\vec x)\; dt$ is a coordinate dependent
statement. It depends on the way one slices up the spacetime with
spacelike hypersurfaces.  Different slicings would lead to different
metrics, and so one has destroyed the coordinate invariance of the
theory right at step one.  This is not a good start for the VSL
programme, as one has performed an act of extreme violence to the
mathematical and logical structure of General Relativistic cosmology,
moving well outside the confines of standard curved-spacetime
Lorentzian geometry.

Another way of viewing this is to start with the
ordinary FLRW metric
\begin{equation}
ds^2 = -c^2 \; dt^2 + a(t)^2 \; h_{ij} \;dx^i \; dx^j,
\end{equation}
and compute the Einstein tensor. In the natural orthonormal basis one
can write
\begin{eqnarray}
G_{\hat t \hat t} 
&=& 
{3\over a(t)^2} \; \left[ {\dot a(t)^2\over c^2}  + K \right] ,
\\
G_{\hat\i \hat\j} 
&=& 
-{\delta_{\hat\i \hat\j} \over a(t)^2} \;
\left[ 
2 {a(t) \; \ddot a(t)\over c^2} + {\dot a(t)^2\over c^2} + K 
\right],
\label{einstein}
\end{eqnarray}
with the spatial curvature $K=0,\pm1$.  If one replaces $c \to c(t)$
{\emph{in the metric}}, then the physics does not change since this
particular ``variable speed of light'' can be undone by a coordinate
transformation: $c \; dt_{\mathrm new} = c(t) \; dt$.  While a
coordinate change of this type will affect the (coordinate) components
of the metric and the (coordinate) components of the Einstein tensor,
the orthonormal components and (by extension) all physical observables
(which are coordinate invariants) will be unaffected.  

An alternative, which does have observable consequences, is the
possibility of replacing $c \to c(t)$ directly {\emph{in the Einstein
tensor}}.  This is the route chosen by Barrow and Magueijo
~\cite{Barrow98a,Barrow98b,Barrow98c}, and by Albrecht
and Magueijo~\cite{Albrecht98,Albrecht99}. Avelino and
Martins~\cite{Avelino-Martins} adopt a slightly different viewpoint,
making the change in the metric, but subject to a time-dependent
redefinition of units.  Then
\begin{eqnarray}
G_{\hat t \hat t}^{\mathrm modified} 
&=&
{3\over a(t)^2} \; \left[ {\dot a(t)^2\over c(t)^2}  + K \right] ,
\\
G_{\hat\i \hat\j}^{\mathrm modified} 
&=&
-{\delta_{\hat\i \hat\j} \over a(t)^2} \;
\left[ 
2 {a(t) \; \ddot a(t)\over c(t)^2} + {\dot a(t)^2\over c(t)^2} + K 
\right].
\end{eqnarray}
%
%----------------------------
Note that the replacement $c \to c(t)$ directly in the Einstein tensor
is a specific implementation of the general prescription presented in 
\cite{Albrecht98}:
``take all time derivatives at fixed $c$ and then replace $c \to c(t)$
in the result''.
%----------------------------

Unfortunately, if one does so, the modified ``Einstein tensor''
so defined is {\emph{not}} covariantly conserved (it does
{\emph{not}} satisfy the contracted Bianchi identities), and this
modified ``Einstein tensor'' is not obtainable from the curvature
tensor of {\emph{any}} spacetime metric.  Indeed, if we define a
timelike vector $V^\mu = (\partial/\partial t)^\mu = (1,0,0,0)$ a
brief computation yields
\begin{equation}
\nabla_\mu G_{\mathrm modified}^{\mu\nu} \propto \dot c(t) \; V^\nu.
\end{equation}
Thus, violations of the Bianchi identities for this modified
``Einstein tensor'' are part and parcel of this particular way of
trying to make the speed of light variable.  Indeed, as we will see
later, in that VSL implementation these violations are the source of
the solution of the flatness problem.  Alternatively one can define
{\emph{modified}} Bianchi identities by moving the RHS above over to
the LHS \cite{Barrow98c} and then speak of these modified Bianchi
identities as being satisfied.  Nevertheless the \emph{usual} Bianchi
identities are violated in their formalism. This may be interpreted as
a statement that such an implementation of VSL is not based on
pseudo-Riemannian geometry (Lorentzian geometry), but that instead one
is dealing with some more complicated structure whose geometric
interpretation is far more complex than usual.

If one couples this modified ``Einstein tensor'' to the stress-energy
via the Einstein equation
\begin{equation}
G_{\mu\nu} = {8\pi \; G_{\mathrm Newton}\over c^4} \; T_{\mu\nu},
\label{gmod}
\end{equation}
then the stress-energy tensor divided by $c^4$ cannot be covariantly
conserved either (here we do not need to specify just yet if we are
talking about a variable $c$ or a fixed $c$), and so $T^{\mu\nu}/c^4$
cannot be variationally obtained from \emph{any} action.
[The factor of $c^4$ is introduced to make sure all the components of
the stress-energy tensor have the dimensions of energy density,
$\varepsilon$  (the same dimensions as pressure, $p$.)  When needed,
mass density will be represented by $\rho$.]
This non-conservation of stress-energy is a tremendous amount of physics
to sacrifice and we do \emph{not} wish to pursue this particular
avenue any further.

Since this point can cause considerable confusion, let us be
clear about what we are claiming: In VSL theories which violate the usual 
Bianchi identities \cite{Albrecht98,Barrow98c}, 
the stress tensor cannot be obtained by variational
differentiation of any local Lagrangian density based on a
pseudo-Riemannian geometry. One can try to generalize the notion of
pseudo-Riemannian geometry but this is an alien procedure from the
standpoint of standard relativity and cosmology.

One of the earliest VSL formulations, and one which does satisfy the 
Bianchi identities, is that of Ellis {\em et al} \cite{EMN}. Inspired by
non-critical string theory, the evolution of $c$ was driven by non-trivial 
renormalization group dynamics associated with the Liouville mode which 
obeys a generalization of the Zamolodchikov C-theorem and therefore 
provides a natural cosmic arrow of time. The advantage of this
formulation is that no extra (and arbitrary) scalar fields are required
to generate the variations in $c$, the disadvantage, as they point out, is
the possibility of making a coordinate transformation to nullify the 
VSL effects.

We feel therefore, that if one wants to uniquely specify that it is
the speed of light that is varying, then the most ``natural'' thing to
do is to seek a theory that contains two natural speed parameters,
call them $\cp$ and $\cg$, and then ask that the ratio of these two
speeds is a time-dependent quantity.  Naturally, once we go beyond
idealized FLRW cosmologies, to include perturbations, we will let this
ratio depend on space as well as time.  Thus we would focus attention
on the dimensionless ratio
\begin{equation}
\zeta \equiv {\cp \over \cg}.
\label{zeta}
\end{equation}
An interesting alternative is to consider the ratio of $\cp$ at
different {\em frequencies}. This ratio is non-trivial in D-brane and
quantum gravity-inspired scenarios \cite{dispersion} which alter the
photon dispersion relation at high energies.

With this idea in mind, we have found that it is simplest to
take $\cg$ to be fixed and position-independent and to set up the
mathematical structure of differential geometry needed in implementing
Einstein gravity: $dx^0 =\cg \; dt$, the Einstein--Hilbert action, the
Einstein tensor,~\etc. One can reserve $\cp$ for photons, and give an
objective meaning to the VSL concept. Observationally, as recently
emphasized by Carlip~\cite{Carlip}, direct experimental evidence tells
us that in the current epoch $\cg\approx\cp$ to within about one
percent tolerance. This limit is perhaps a little more relaxed than
one would have naively expected, but the looseness of this limit is a
reflection of the fact that direct tests of General Relativity are
difficult due to the weakness of the gravitational coupling
$G_{\mathrm Newton}$. 

Although we will focus on models and systems of units in which $\cp$
varies while $\cg$ is fixed, in Appendix~\ref{cgvary} we consider the
reverse. This is important for discussions of varying fine-structure
constant $\alpha$. Since $\alpha \propto \cp^{-1}$, the models we
present in the following sections {\em do lead to variation of the
  fine-structure constant}.  This issue will be important in
model-building if the Webb {\etal} \cite{varyalpha} results on
time-varying $\alpha$ are confirmed.

The above approach naturally leads us into the
realm of two-metric theories, and the next section will be devoted to
discussing the origin of our proposal. In brief, we will advocate
using at least {\emph{two}} metrics: a spacetime metric
$g_{\alpha\beta}$ describing gravity, and a second ``effective
metric'' $[g^{\mathrm em}]_{\alpha\beta}$ describing the propagation
of photons. Other particle species could, depending on the specific
details of the model we envisage, couple either to their own
``effective metric'', to $g$, or to $g^{\mathrm em}$.

Specific early examples of a VSL model based on a two-metric theory
are those of Moffat ~\cite{Moffat93a,Moffat93b}, with a more recent
implementation being that of Drummond \cite{Drummond}.  Moffat chooses
to keep $\cp$ fixed and let $\cg$ vary, which leads to some
translation difficulties in comparing those papers with the current
one; but it is clear that there are substantial areas of
agreement. This paper can be viewed as an extension of those previous
investigations.

To help set the background, we wish to emphasize that the basic idea
of a quantum-induced effective metric, which affects only photons and
differs from the gravitational metric, is actually far from
radical. This concept has gained a central role in the discussion of
the propagation of photons in non-linear electrodynamics. In
particular, we stress that ``anomalous'' ($\cp > \cg$) photon speeds
have been calculated in relation with the propagation of light in the
Casimir vacuum~\cite{Sch90,Bar90,SchBar93}, as well as in
gravitational fields~\cite{DH80,DS94,DS96,Shore96}. 

These articles
have shown that special quantum vacuum states (associated with
``polarization'' of the vacuum) can lead to a widening of lightcones
(although possibly only in some directions and for special photon
polarization). In recent papers~\cite{DG98,Novello99} it has been
stressed that such behaviour can be described in a geometrical way by
the introduction of an effective metric which is related to the
spacetime metric and the renormalized stress-energy tensor by a
relation such as
\begin{equation}
[\gem^{-1}]^{\mu\nu}=
A\; g^{\mu\nu} + B\; \langle\psi| T^{\mu\nu} |\psi\rangle, 
\end{equation}
where $A$ and $B$ depend on the detailed form of the
effective (one-loop) Lagrangian for the electromagnetic field.

\bigskip
\underline{Warning:} 
We will always raise and lower indices using the spacetime metric
$g$. This has the side-effect that one can no longer use index
placement to distinguish the matrix $[\gem]$ from its matrix inverse
$[\gem^{-1}]$.  (Since $[\gem]^{\mu\nu} \equiv g^{\mu\sigma} \;
g^{\nu\rho} \; [\gem]_{\sigma\rho} \neq [\gem^{-1}]^{\mu\nu}.$)
Accordingly, whenever we deal with the EM metric, we will always
explicitly distinguish $[\gem]$ from its matrix inverse $[\gem^{-1}]$.
\bigskip

It is important to note that such effects can safely be described
without needing to take the gravitational back reaction into
account. The spacetime metric $g$ is only minimally affected by the
vacuum polarization, because the formula determining $[g^{\mathrm
em}]$ is governed by the fine structure constant, while backreaction
on the geometry is regulated by Newton's constant.  Although these
deviations from standard propagation are extremely tiny for the above
quoted cases (black holes and the Casimir vacuum) we can ask ourselves
if a similar sort of physics could have been important in the early
evolution of our universe.

Drummond and Hathrell \cite{DH80} have, for example, computed one-loop
vacuum polarization corrections to QED in the presence of a
gravitational field.  They show that at low momenta the effective
Lagrangian is
%----------------------------------------------------------------
\begin{eqnarray}
{\cal L} &=& 
-\frac{1}{4} \; F_{\mu\nu} \; F^{\mu \nu} 
\nonumber\\
&&
- \frac{1}{4m_e^2} (\beta_1 \; R \; F_{\mu\nu} \; F^{\mu \nu} 
+  \beta_2 \; R_{\mu\nu} \; F^{\mu \alpha} \; F^{\nu}{}_{\alpha})
\nonumber\\
&& 
- \frac{\beta_3}{4m_e^2} \; R_{\mu\nu\alpha\beta} \; F^{\mu \alpha} \;
F^{\mu
\beta}\,.
\label{dh}
\end{eqnarray}
Drummond and Hathrell were able to compute the low momentum
coefficients $\beta_i, i = 1\dots3$, but their results are probably
not applicable to the case $R/m_e \gg 1$ of primary interest here. It
is the qualitative structure of their results that should be compared
with our prescriptions as developed in the next section.
 
In the main body of this paper we sketch out a number of scenarios
based on two-metric interpretations of the VSL idea. We present
different models that are consistent (\ie, mathematically and
logically consistent), and which satisfy zeroth-order
compatibility with observations (\ie, at least reduce to ordinary
special relativity in the here and now).  We also indicate how the
various puzzles of the standard cosmological model can be formulated
in this language, and start a preliminary analysis of these issues.

Since doing anything to damage and violate Lorentz symmetry is at
first glance a rather radical step, we also wish to add a few words
regarding the various approaches to the breaking of Lorentz invariance
that are well-established in the literature. Perhaps the most
important observation is that quantum field theories that are not
Lorentz invariant can nevertheless exhibit an approximate Lorentz
invariance in the low energy limit.  See, for instance, the work of
Nielsen \etal \cite{Nielsen78,Nielsen83a,Nielsen83b}, where they
demonstrate that Lorentz invariance is often a stable infrared fixed
point of the renormalization group flow of a quantum field theory.
An alternative model for the breakdown of Lorentz invariance has also
been discussed by Everett \cite{Everett76a,Everett76b}.

Additionally, there are physical systems (in no sense relativistic,
and based on the flowing fluid analogy for Lorentzian spacetimes) that
demonstrate that Lorentz invariance can arise as a low energy property
\cite{Unruh81,Jacobson91,Jacobson93,Visser93,Unruh94,Visser98}. In the
flowing fluid analogy for Lorentzian spacetimes the fluid obeys the
non-relativistic Euler and continuity equations, while sound waves
propagating in the fluid behave as though they ``feel'' a Lorentzian
metric (with appropriate symmetries) that is built algebraically out
of the dynamical variables describing the fluid flow.

Furthermore, as yet another example of ``soft'' Lorentz symmetry
breaking we mention the well-studied Scharnhorst
effect~\cite{Sch90,Bar90,SchBar93}, wherein quantum vacuum effects
lead to an anomalous speed of light for photons propagating
perpendicular to a pair of conducting metal plates. The relevant
one-loop quantum physics is neatly summarized by the Euler--Heisenberg
effective Lagrangian, which explicitly exhibits a symmetry under the
full (3+1)-dimensional Lorentz group. However the ground state (field
theoretic vacuum state) exhibits a {\em reduced} symmetry, being
invariant only under boosts that are parallel to the plates. In this
situation the boundary conditions have ``softly'' broken the symmetry
from (3+1)-dimensional Lorentz invariance down to (2+1)-dimensional
Lorentz invariance, even though the fundamental physics encoded in the
bulk Lagrangian is still manifestly symmetric under the larger group.

These comments bolster the view that we should not be too worried by a
gentle breaking of Lorentzian symmetry. In this vein, Coleman and
Glashow have recently investigated the possibility of small,
renormalizable perturbations to the standard model which break Lorentz
invariance while preserving the anomaly cancellation \cite{CG}.  These
perturbations are important at high energies and may provide an
explanation for the existence of ultra-high energy cosmic rays beyond
the GZK cut-off \cite{CG}.

Finally, we should again remind the reader that VSL implementations
based on two-metric theories are certainly closer in spirit to the
approaches of Moffat \& Clayton
\cite{Moffat93a,Moffat93b,Moffat98,Clayton-Moffat-1,Clayton-Moffat-2}
and Drummond \cite{Drummond}, than to the early
Albrecht--Barrow--Magueijo
\cite{Albrecht98,Barrow98c,Albrecht99} and
Avelino--Martins~\cite{Avelino-Martins} prescriptions.  We have so far
been unable to develop any really clean geometrodynamic framework that
more closely parallels the phenomenological approach of the Barrow
{\etal} approach, though we hope to be able to return to that issue in
the future.

In Table 1 we give a list of variables and symbols used in this paper
together with a brief description and appropriate defining equation.

%---------------------------------------------------------------------
\section{Two-metric VSL cosmologies} \label{2metric}
%----------------------------------------------------------------------

Based on the preceding discussion, we think that the first step
towards making a ``geometric'' VSL cosmology is to write
a two-metric theory in the form
\begin{eqnarray}
S_{I} &=& 
\int d^4 x \sqrt{-g} \; 
\left\{ R(g) + \L_{\mathrm matter}(g) \right\}
\nonumber\\
&+&
\int d^4 x \sqrt{-\gem} \; 
\left\{
[\gem^{-1}]^{\alpha\beta}  \; F_{\beta\gamma} \;
[\gem^{-1}]^{\gamma\delta} \; F_{\delta\alpha}
\right\}.
\label{s1}
\end{eqnarray}
We have made the first of many \emph{choices} here by choosing the
volume element for the electromagnetic Lagrangian to be $\sqrt{-\gem}$
rather than, say $\sqrt{-g}$. This has been done to do minimal damage
to the electromagnetic sector of the theory. As long as we confine
ourselves to making \emph{only} electromagnetic measurements this
theory is completely equivalent to ordinary curved space
electromagnetism in the spacetime described by the metric $\gem$. As
long as we \emph{only} look at the ``matter'' fields it is only the
``gravity metric'' $g$ that is relevant.

Since the photons couple to a second, separate metric, distinct from
the spacetime metric that describes the gravitational field, we can
now give a precise physical meaning to VSL. If the two null-cones
(defined by $g$ and $\gem$, respectively) do not coincide one has a
VSL cosmology. Gravitons and all matter except for photons, couple to
$g$. Photons couple to the electromagnetic metric $\gem$.  A more
subtle model is provided by coupling all the gauge bosons to $\gem$,
but everything else to $g$.
\begin{eqnarray}
S_{II} &=& 
\int d^4 x \sqrt{-g} \; 
\left\{ R(g) + \L_{\mathrm fermions}(g,\psi) \right\} 
\nonumber\\
&+&
\int d^4 x \sqrt{-\gem} \; 
\Tr \left\{
[\gem^{-1}]^{\alpha\beta}  \; F_{\beta\gamma}^{\mathrm gauge} \;
[\gem^{-1}]^{\gamma\delta} \; F_{\delta\alpha}^{\mathrm gauge} 
\right\}.
\nonumber\\
&&
\label{s2}
\end{eqnarray}
For yet a third possibility: couple \emph{all} the matter fields to
$\gem$, keeping gravity as the only field coupled to $g$.
That is
\begin{eqnarray}
S_{III} &=& 
\int d^4 x \sqrt{-g} \;  R(g)
\nonumber\\
&+&
\int d^4 x \sqrt{-\gem} \; 
\left\{ \L_{\mathrm fermions}(\gem,\psi) \right\} 
\nonumber\\
&+&
\int d^4 x \sqrt{-\gem} \; 
\Tr \left\{
[\gem^{-1}]^{\alpha\beta}  \; F_{\beta\gamma}^{\mathrm gauge} \;
[\gem^{-1}]^{\gamma\delta} \; F_{\delta\alpha}^{\mathrm gauge}
\right\}.
\nonumber\\
&&
\end{eqnarray}
Note that we have used $dx^0 = c \; dt$, with the $c$ in question
being $\cg$. It is this $\cg$ that should be considered fundamental,
as it appears in the local Lorentz transformations that are the
symmetry group of all the non-electromagnetic interactions. It is just
that $\cg$ is no longer the speed of ``light''.

Most of the following discussion will focus on the first model $S_I$,
but it is important to realize that VSL cosmologies can be implemented
in many different ways, of which the models I, II, and III are the
cleanest exemplars.  We will see later that there are good reasons to
suspect that model III is more plausible than models I or II, but we
concentrate on model I for its pedagogical clarity. If one wants a
model with even more complexity, one could give a \emph{different}
effective metric to each particle species. A model of this type would
be so unwieldy as to be almost useless.

If there is no relationship connecting the EM metric to the gravity
metric, then the theory has too much freedom to be useful, and the
equations of motion are under-determined. To have a useful theory we
need to postulate some relationship between $g$ and $\gem$, which in
the interest of simplicity we take to be algebraic.  A particularly
simple electromagnetic (EM) metric we have found useful to consider
is\footnote{
%---------------------------------------------------------------------
The form of this metric is similar to the Kerr--Schild--Trautmann
ansatz for generating exact solutions: $g_{ab}=\eta_{ab}-2V k_a k_b$,
where $k_a$ is null in both the flat and non-flat metrics. $k_a$ is
geodesic if and only if $T_{ab}k^a k^b=0$. This generates a 
family of vacuum and
Einstein--Maxwell solutions \cite{exact}.}
%----------------------------------------------------------------------
%
\begin{equation}
[\gem]_{\alpha\beta} = 
g_{\alpha\beta} - 
(A\; M^{-4}) \;
\nabla_\alpha \chi \; \nabla_\beta \chi,
\label{chimodel}
\end{equation}
with the inverse metric
\begin{equation}
[\gem^{-1}]^{\alpha\beta} = 
g^{\alpha\beta} + 
(A\; M^{-4}) \;
{
\nabla^\alpha \chi \; \nabla^\beta \chi
\over
1 + (A\; M^{-4}) \;
(\nabla^\alpha \chi)^2
}.
\end{equation}
Here we have introduced a dimensionless coupling $A$ and taken
$\hbar=\cg=1$, in order to give the scalar field $\chi$ its canonical
dimensions of mass-energy.\footnote{
%--------------------------------------------
Remember that indices are always raised and/or lowered by using the
gravity metric $g$. Similarly, contractions always use the
gravity metric $g$. If we ever need to use the EM metric to contract
indices we will exhibit it explicitly.}
%--------------------------------------------
The normalization energy scale, $M$, is defined in terms of $\hbar$,
$G_{\mathrm Newton}$, and $\cg$.  The EM lightcones can be much wider
than the standard (gravity) ones without inducing a large backreaction
on the spacetime geometry from the scalar field $\chi$, provided $M$
satisfies $M_{\mathrm Electroweak}<M<M_{\mathrm Pl}$. The presence of
this dimensionfull coupling constant implies that when viewed as a
quantum field theory, {\VSL} cosmologies will be non-renormalizable.
In this sense the energy scale $M$ is the energy at which the
non-renormalizability of the $\chi$ field becomes important. (This is
analogous to the Fermi scale in the Fermi model for weak interactions,
although in our case $M$ could be as high as the GUT scale).  Thus,
{\VSL} models should be viewed as ``effective field theories'' valid
for sub-$M$ energies. In this regard {\VSL} models are certainly no
worse behaved than many of the models of cosmological inflation and/or
particle physics currently extant.

In comparison, note that Moffat~\cite{Clayton-Moffat-1} introduces a
somewhat similar vector-based model for an effective metric which in
our notation would be written as
\begin{equation}
[\gem]_{\alpha\beta} = 
g_{\alpha\beta} - 
(A\; M^{-2}) \;
V_\alpha \; V_\beta,
\end{equation}
with the inverse metric
\begin{equation}
[\gem^{-1}]^{\alpha\beta} = 
g^{\alpha\beta} + 
(A\; M^{-2}) \;
{
V^\alpha \; V^\beta
\over
1 + (A\; M^{-2}) \;(V^\alpha)^2
}.
\end{equation}
However there are many technical differences between that paper and
this one, as will shortly become clear. In the more recent
paper~\cite{Clayton-Moffat-2} a scalar-based scenario more similar to
our own is discussed.

The evolution of the scalar field $\chi$ will be assumed to be
governed by some VSL action
\begin{equation}
S_{\mathrm VSL} =   \int d^4 x \sqrt{-g} \; \L_{\mathrm VSL}(\chi).
\end{equation}
We can then write the complete action for model I as
\begin{eqnarray}
S_{I} &=& 
\int d^4 x \sqrt{-g} \; \left\{ R(g) + \L_{\mathrm matter} \right\} 
\nonumber\\
&+&
\int d^4 x \sqrt{-\gem(\chi)} \;
\left\{
[\gem^{-1}]^{\alpha\beta}(\chi)  \; F_{\beta\gamma} \;
[\gem^{-1}]^{\gamma\delta}(\chi) \; F_{\delta\alpha}
\right\} 
\nonumber\\
&+&
\int d^4 x \sqrt{-g} \; \L_{\mathrm VSL}(\chi)
\nonumber\\
&+& \int d^4 x \sqrt{-g} \; \L_{\mathrm NR}(\chi,\psi),
\end{eqnarray}
where $\L_{\mathrm NR}(\chi,\psi)$ denotes the non-renormalizable
interactions of $\chi$ with the standard model.

Let us suppose the potential in this VSL action has a global minimum,
but the $\chi$ field is displaced from this minimum in the early
universe: either trapped in a metastable state by high-temperature
effects or displaced due to chaotic initial conditions. The transition
to the global minimum may be either of first or second order and
during it $\nabla_\alpha \chi \neq0$, so that $\gem \neq g$. Once the
true global minimum is achieved, $\gem = g$ again. Since one can
arrange $\chi$ today to have settled to the true global minimum,
current laboratory experiments would automatically give $\gem=g$.

It is only via observational cosmology, with the possibility of
observing the region where $\gem \neq g$ that we would expect VSL
effects to manifest themselves.  We will assume the variation of the
speed of light to be confined to very early times, of order of the GUT
scale, and hence none of the low-redshift physics can be directly
affected directly by this transition. We will see in section
\ref{obst} how indirect tests for the presence of the $\chi$ field are
indeed possible.

%-----------------------------------------------------
\begin{figure*}[t]
\epsfxsize=3.4in
\epsffile{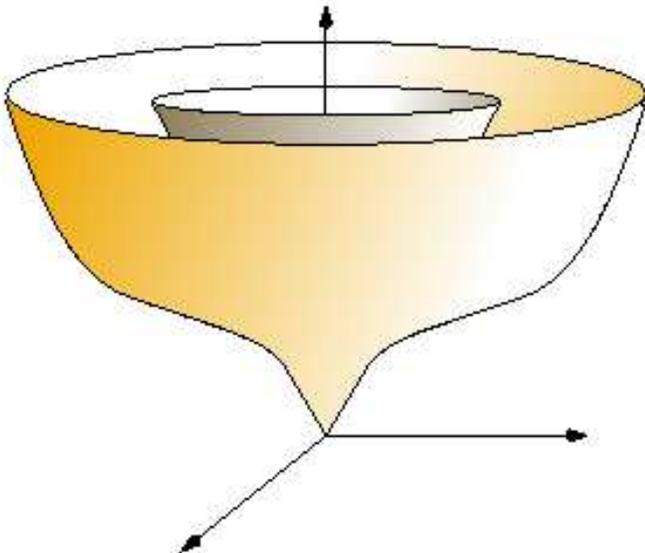}
\caption{
A schematic illustration of the two future null cones
$C^+_{\mathrm{gravity}}$ and $C^+_{\mathrm{photon}}$. Initially they
coincide, followed by a transition after which $c_{\mathrm{photon}}
\gg c_{\mathrm{gravity}}$ and then by another transition in which
$c_{\mathrm{photon}} \simeq c_{\mathrm{gravity}}$. }
\end{figure*}
%-----------------------------------------------------

Note that in the metastable minimum $V(\chi)\geq 0$, thus the scalar
field $\chi$ can mimic a cosmological constant, as long as the kinetic
terms of the VSL action are negligible when compared to the potential
contribution.  If the lifetime of the metastable state is too long, a
de Sitter phase of exponential expansion will ensue. Thus, the VSL
scalar has the possibility of driving an inflationary phase in its own
right, over and above anything it does to the causal structure of the
spacetime (by modifying the speed of light). While this direct
connection between VSL and inflation is certainly interesting in its
own right, we prefer to stress the more interesting possibility that,
by coupling an independent inflaton field $\phi$ to $\gem$, $\chi$VSL
models can be used to improve the inflationary framework by enhancing
its ability to solve the cosmological puzzles. We will discuss this
issue in detail in section \ref{hete}.

During the transition, (adopting FLRW coordinates on the spacetime),
we see
\begin{equation}
[\gem]_{tt} 
= -1 - (A \; M^{-4}) \; (\partial_t \chi)^2 
\leq -1.
\end{equation}
This means that the speed of light for photons will be larger than the
``speed of light'' for everything else --- the photon null cone will
be wider than the null cone for all other forms of matter.\footnote{
%-----------------------------------------------------
For other massless fields the situation depends on whether we use
model I, II, or III. In model I it is \emph{only} the photon that sees
the anomalous light cones, and neutrinos for example are
unaffected. In model II all gauge bosons (photons, $W^\pm$, $Z^0$, and
gluons) see the anomalous light cones. Finally, in model III
everything \emph{except} gravity sees the anomalous light cones.}
%---------------------------------------------------
Actually one has
\begin{equation}
\cp^2 = 
\cg^2 
\left[1 + (A \; M^{-4}) \; (\partial_t \chi)^2\right] 
\geq 
\cg^2 .
\label{cphoton}
\end{equation}
The fact that the photon null cone is wider implies that ``causal
contact'' occurs over a larger region than one thought it did --- and
this is what helps smear out inhomogeneities and solve the horizon
problem.

The most useful feature of this model is that it gives a precise
\emph{geometrical} meaning to VSL cosmologies: something that is
difficult to discern in the extant literature.

Note that this model is by no means unique: (1) the VSL potential is
freely specifiable, (2) one could try to do similar things to the
Fermi fields and/or the non-Abelian gauge fields --- use one metric
for gravity and $\gem$ for the other fields.  We wish to emphasize
some features and pitfalls of two-metric VSL cosmologies:

$\bullet$ The causal structure of spacetime is now ``divorced'' from
the null geodesics of the metric $g$.  Signals (in the form of
photons) can travel at a speed $\cp \geq \cg$.

$\bullet$ We must be extremely careful whenever we need to assign a
specific meaning to the symbol $c$. We are working with a
\emph{variable\,} $\cp$, which has a larger value than the standard
one, and a \emph{constant\,} $\cg$ which describes the speed of
propagation of all the other massless particles. In considering the
cosmological puzzles and other features of our theory (including the
``standard'' physics) we will always have to specify if the quantities
we are dealing with depend on $\cp$ or $\cg$.

$\bullet$ Stable causality: If the gravity metric $g$ is causally
stable, if the coupling $A \geq 0$, and if $\partial_\mu \chi$ is a
timelike vector with respect to the gravity metric, then the photon
metric is also causally stable. This eliminates the risk of nasty
causal problems like closed timelike loops. This observation is
important since with two metrics (and two sets of null cones), one
must be careful to not introduce causality violations --- and if the
two sets of null cones are completely free to tip over with respect to
each other it is very easy to generate causality paradoxes in the
theory.

$\bullet$ If $\chi$ is displaced from its global minimum we expect it
to oscillate around this minimum, causing $\cp$ to have periodic
oscillations. This would lead to dynamics very similar to that of
preheating in inflationary scenarios~\cite{Preheating}.

$\bullet$ During the phase in which $c_{\mathrm{photon}} \gg
c_{\mathrm{gravity}}$ one would expect photons to emit gravitons in an
analogue of the Cherenkov radiation.  We will call this effect
\emph{Gravitational Cherenkov Radiation}.  This will cause the
frequency of photons to decrease and will give rise to an additional
stochastic background of gravitons.

$\bullet$ Other particles moving faster than $c_{\mathrm{gravity}}$
(\ie, models II and III) would slow down and become subluminal
relative to $c_{\mathrm{gravity}}$ on a characteristic time-scale
associated to the emission rate of gravitons.  There will therefore be
a natural mechanism for slowing down massive particles to below
$c_{\mathrm{gravity}}$.

$\bullet$ In analogy to photon Cherenkov emission \cite{cherenkov},
longitudinal graviton modes may be excited due to the non-vacuum
background \cite{vEE}.

%-------------------------------------------------------------------
\section{Stress-energy tensor, equation of state, and equations of motion}
%-------------------------------------------------------------------
\subsection{The two stress-energy tensors}
%-------------------------------------------------------------------

The definition of the stress-energy tensor in a VSL cosmology is
somewhat subtle since there are two distinct ways in which one could
think of constructing it. If one takes gravity as being the primary
interaction, it is natural to define
\begin{equation}
T^{\mu\nu} = {2\over\sqrt{-g}} {\delta S\over\delta g_{\mu\nu}},
\end{equation}
where the metric variation has been defined with respect to the
gravity metric. This stress-energy tensor is the one that most
naturally shows up in the Einstein equation. One could also think of
defining a different stress-energy tensor for the photon field (or in
fact any form of matter that couples to the photon metric) by varying
with respect to the photon metric, that is
\begin{equation}
\tilde T^{\mu\nu} = {2\over\sqrt{-g_{\mathrm em}}}
{\delta S\over\delta g^{\mathrm em}_{\mu\nu}}.
\end{equation}
This definition is most natural when one is interested in
non-gravitational features of the physics.

In the formalism we have set up, by using the chain rule and the
relationship that we have assumed between $g_{\mathrm em}$ and $g$, it
is easy to see that
\begin{equation}
\label{E:two-stress-tensors}
T_{\mathrm em}^{\mu\nu} = 
\sqrt{g_{\mathrm em}\over g} \; \tilde T_{\mathrm em}^{\mu\nu} = 
\sqrt{1-(A \; M^{-4})[(\nabla^\alpha \chi)^2]} \; \;
\tilde T_{\mathrm em}^{\mu\nu}.
\end{equation}
Thus, these two stress-energy tensors are very closely related. When
considering the way the photons couple to gravity, the use of
$T_{\mathrm em}^{\mu\nu}$ is strongly recommended. Note that
$T_{\mathrm em}^{\mu\nu}$ is covariantly conserved with respect to
$\nabla_g$, whereas $\tilde T_{\mathrm em}^{\mu\nu}$ is conserved with
respect to $\nabla_{\gem}$. It should be noted that $\tilde T_{\mathrm
em}^{\mu\nu}$ is most useful when discussing the non-gravitational
behaviour of matter that couple to $\gem$ rather than $g$. (Thus in
type I models this means we should only use it for photons.)  For
matter that couples to $g$ (rather than to $\gem$), we have not found
it to be indispensable, or even useful, and wish to discourage its use
on the grounds that it is dangerously confusing.

An explicit calculation, assuming for definiteness a type I model and
restricting attention to the electromagnetic field, yields
\begin{eqnarray}
T_{\mathrm em}^{\mu\nu} &=& 
\sqrt{1-(A \; M^{-4})[(\nabla^\alpha\chi)^2]}
\nonumber\\
&&
\times
\Bigg\{ 
[\gem^{-1}]^{\mu\sigma} \; F_{\sigma\rho} 
\; [\gem^{-1}]^{\rho\lambda} 
\; F_{\lambda\pi} \; [\gem^{-1}]^{\pi\nu}
\nonumber\\
&&
\qquad
- {1\over4} [\gem^{-1}]^{\mu\nu} \; (F^2)
\Bigg\},
\end{eqnarray}
with
\begin{eqnarray}
(F^2) = 
[\gem^{-1}]^{\alpha\beta}  \; F_{\beta\gamma} \;
[\gem^{-1}]^{\gamma\delta} \; F_{\delta\alpha}.
\end{eqnarray} 
(In particular, note that both $\tilde T_{\mathrm em}^{\mu\nu}$ and
$T_{\mathrm em}^{\mu\nu}$ are traceless with respect to $\gem$, not
with respect to $g$. This observation will prove to be very useful.)

%-------------------------------------------------------------------
\subsection{Energy density and pressure: the photon equation--of--state}
%-------------------------------------------------------------------

In an FLRW universe the high degree of symmetry implies that the
stress-energy tensor is completely defined in terms of energy density
and pressure. We will define \emph{the} physical energy density and
pressure as the appropriate components of the stress-energy tensor
when referred to an orthonormal basis \emph{of the metric that enters
the Einstein equation} (from here on denoted by single-hatted indices)
\begin{eqnarray}
\varepsilon &=&  
 T^{{\hat t} {\hat t}} 
= 
T^{tt}/|g^{tt}| 
= 
|g_{tt}|\; T^{tt},
\label{energyden}\\
p  &=&  
{1\over 3} \delta_{{\hat\i} {\hat\j}} \;  T^{{\hat\i} {\hat\j}} = 
{1\over3} \; g_{ij} \; T^{ij}.
\label{E:pressure}
\end{eqnarray}
It is this $\varepsilon$ and this $p$ that will enter the Friedmann
equations governing the expansion and evolution of the universe.

On the other hand, if one defines the stress-energy tensor in terms of
a variational derivative with respect to the electromagnetic metric,
then when viewed from an orthonormal frame adapted to the
{\emph{electromagnetic}} metric (denoted by double hats), one will
naturally define {\emph{different}} quantities for the energy density
$\tilde\varepsilon$ and pressure $\tilde p$. We can then write
\begin{eqnarray}
\tilde\varepsilon &=&  
\tilde T^{\hat{\hat t} \hat{\hat t}} 
= 
\tilde T^{tt}/|g^{tt}_{\mathrm em}| 
= 
|g_{tt}^{\mathrm em}|\; \tilde T^{tt},
\\
\tilde p  &=&  
{1\over 3} \delta_{\hat{\hat\i} \hat{\hat\j}} \; 
           \tilde T^{\hat{\hat\i} \hat{\hat\j}} = 
{1\over3} \; g_{ij}^{\mathrm em} \; \tilde T^{ij}.
\end{eqnarray}
{From} our previous discussion [equation (\ref{E:two-stress-tensors})]
we know that the two definitions of stress-energy are related, and
using the symmetry of the FLRW geometry we can write
\begin{equation}
T^{\mu\nu} = {\cp\over\cg} \; \tilde T^{\mu\nu}.
\label{E:two-stress-tensors-FLRW}
\end{equation}
If we combine this equation with the previous definitions, we have
\begin{eqnarray}
\varepsilon &=& {\cg\over\cp} \; \tilde\varepsilon,
\\
p &=& {\cp\over\cg} \; \tilde p.
\label{E:p}
\end{eqnarray}
(Note that the prefactors are \emph{reciprocals} of each other.)  From
a gravitational point of view any matter that couples to the photon
metric has its energy density depressed and its pressure enhanced by a
factor of $\cg/\cp$ relative to the energy density and pressure
determined by ``electromagnetic means''.  This ``leverage'' will
subsequently be seen to have implications for SEC violations and
inflation.

In order to investigate the equation of state for the photon field,
our starting point will be the standard result that the stress-energy
tensor of photons is traceless. By making use of the tracelessness and
symmetry arguments one can (in one-metric theories) deduce the
relationship between the energy density and the pressure $\varepsilon
= 3 p$.  However, in two-metric theories (of the type presented here)
the photon stress-energy tensor is traceless with respect to
$g_{\mathrm em}$, but not with respect to $g$.  Thus in this bi-metric
theory we have
\begin{equation}
\tilde \varepsilon = 3 \tilde p.
\end{equation}
When translated into $\varepsilon$ and $p$, (quantities that will
enter the Friedmann equations governing the expansion and evolution of
the universe), this implies
\begin{equation}
p_{\mathrm photons} = 
{1\over3} \; \varepsilon_{\mathrm photons} \; {\cp^2\over\cn^2}.
\label{eosp}
\end{equation}

As a final remark it is interesting to consider the speed of sound
encoded in the photon equation of state. If we use the relationship
$\rho_{\mathrm photons} = \varepsilon_{\mathrm photons}/ \cg^2$, we
can write
\begin{equation}
\rho_{\mathrm photons}= 
{3 \; p_{\mathrm photons}\over\cp^2}.
\label{rho}
\end{equation}
And therefore 
\begin{equation}
(c_{\mathrm sound})_{\mathrm photons} = 
\sqrt{\partial p_{\mathrm photons}\over\partial\rho_{\mathrm photons}} = 
{\cp\over\sqrt{3}}.
\end{equation}
That is, oscillations in the density of the photon fluid propagate at
a relativistic speed of sound which is $1/\sqrt{3}$ times the speed of
``light'' \emph{as seen by the photons}.

More generally, for highly relativistic particles we expect
\begin{equation}
\varepsilon_i = 
3 \; p_i \; {\cg^2\over c_i^2},
\end{equation}
and
\begin{equation}
(c_{\mathrm sound})_i = 
{c_i\over\sqrt{3}}.
\end{equation}
Note that we could define the mass density (as measured by
electromagnetic means) in terms of $\tilde\rho_{\mathrm photons} =
\tilde\varepsilon_{\mathrm photons}/\cp^2$.  This definition yields
the following identity
\begin{equation}
\rho_{\mathrm photons} = {\cp\over \cn} \;\tilde \rho_{\mathrm photons}.
\end{equation}
If the speed of sound is now calculated in terms of $\tilde p_{\mathrm
photons}$ and $\tilde \rho_{\mathrm photons}$ we get the same result
as above.

%-------------------------------------------------------------------
\subsection{Equations of motion}
%-------------------------------------------------------------------

The general equations of motion based on model I can be written as
\begin{equation}
G_{\mu\nu} = 
{8\pi \; G_{\mathrm Newton} \over \cg^4} \;
\left(
T_{\mu\nu}^{\mathrm VSL} + 
T_{\mu\nu}^{\mathrm em} + 
T_{\mu\nu}^{\mathrm matter} 
\right).
\end{equation}
All of these stress-energy tensors have been defined with the
``gravity prescription''
\begin{equation}
T_i^{\mu\nu} = {2\over\sqrt{-g}} {\delta S_i\over\delta g_{\mu\nu}}.
\end{equation}

In a FLRW spacetime the Friedmann equations (summing
over all particles present) for a {\VSL} cosmology read as follows
\begin{eqnarray}
{\left({\dot a\over a}\right)}^2
&=&
{8\pi G\over3\cg^2} \; \sum_{i} \varepsilon_{i}
-{K \cg^2\over a^2},
\label{f1}
\\
{\ddot a\over a}
&=&
-{4\pi G\over 3\cg^2}\; \sum_{i}{\left(\varepsilon_{i}+3p_{i}\right)}.
\label{f2}
\end{eqnarray}
where, as usual, $K=0,\pm1$.

The constant ``geometric'' speed of light implies that we get from the
Friedmann equation separate conservation equations valid for each
species individually (provided, as is usually assumed for at least
certain portions of the universe's history, that there is no
significant energy exchange between species)
\begin{equation}
\dot\varepsilon_i +
3{\dot a\over a}{\left(\varepsilon_{i}+{p_{i}}\right)}=0.
\end{equation}
In the relativistic limit we have already seen, from equation
(\ref{eosp}), that $p_{i}= {1\over3} \varepsilon_{i} \;
(c_i^2/\cg^2)$.  [We are generalizing slightly to allow each
particle species to possess its own ``speed-of-light''.] So we can
conclude that
\begin{equation}
\dot\varepsilon_i +
\left(3+ {c_i^2\over \cg^2}\right) {\dot a\over a} \; \varepsilon_{i}=0.
\end{equation}
Provided $c_i$ is slowly changing with respect to the expansion of the
universe (and it is not at all clear whether such an epoch ever
exists), we can write for each relativistic species
\begin{equation}
\varepsilon_i \; a^{3+ (c_i^2/\cg^2)} \approx {\mathrm constant}.
\end{equation}
This is the generalization of the usual equation $(\varepsilon_i a^4
\approx {\mathrm constant})$ for relativistic particles in a
constant-speed-of-light model. This implies that energy densities will
fall much more rapidly than naively expected in this bi-metric VSL
formalism, provided $c_i > \cg$.

%-------------------------------------------------------------------
\section{Cosmological puzzles and Primordial Seeds}
%-------------------------------------------------------------------

In the following we will discuss the main cosmological puzzles
showing how they are mitigated (if not completely solved) by the
\VSL~models.  Given its complexity, the peculiar case of the flatness
problem will be treated in a separate section.

%--------------------------------------------------
\subsection{The isotropy and horizon problems}
%--------------------------------------------------

One of the major puzzles of the standard cosmological model is that
the isotropy of the CMB seems in conflict with the best estimates of
the size of causal contact at last scattering. The formula for the
(coordinate) size of the particle horizon at the time of last
scattering $t_*$ is
\begin{equation}
R_{\mathrm particle-horizon}(t_*) = \int_0^{t_*} {\cg \; dt \over a(t)}.
\end{equation}
For photons this should now be modified to
\begin{eqnarray}
R_{\mathrm photon-horizon}(t_*) 
&=& \int_0^{t_*} {\cp \; dt \over a(t)} 
\\
&\geq& R_{\mathrm particle-horizon}(t_*).
\end{eqnarray}
The quantity $R_{\mathrm photon-horizon}$ sets the distance scale over
which photons can transport energy and thermalize the primordial
fireball.  On the other hand, the coordinate distance to the surface
of last scattering is
\begin{equation}
R_{\mathrm last-scattering}(t_*,t_0) = \int_{t_*}^{t_0} {\cp \; dt \over a(t)}.
\end{equation}
(Here $t_0$ denotes the present epoch.)  The observed large-scale
homogeneity of the CMB implies (since you want the CMB coming at you
from opposite points on the sky to be the same without any artificial
fine-tuning)
\begin{equation}
R_{\mathrm photon-horizon}(t_*) \geq 2 \; R_{\mathrm last-scattering}(t_*,t_0),
\end{equation}
which can be achieved by having $\cp \gg \cg$ early in the expansion.
(In order not to change late-time cosmology too much it is reasonable
to expect $\cp \approx \cg$ between last scattering and the present
epoch.) Instead of viewing our observable universe as an inflated
small portion of the early universe (standard inflationary cosmology),
we can say that in a VSL framework the region of early causal contact
is underestimated by a factor that is roughly approximated by the
ratio of the maximum photon speed to the speed with which
gravitational perturbations propagate.

We can rephrase the horizon problem as a constraint on the ratio
between the photon horizon at last scattering and the photon horizon
at the present day. Indeed if we add $2 R_{\mathrm
photon-horizon}(t_*)$ to both sides of the previous equation, then
\begin{equation}
3 \; R_{\mathrm photon-horizon}(t_*) \geq 2 \; R_{\mathrm photon-horizon}(t_0).
\end{equation}
In terms of the {\em physical distance} to the photon horizon
($\ell(t) = a(t) R(r)$), this implies
\begin{equation}
\label{E:horizon}
\ell_{\mathrm photon-horizon}(t_*) 
\geq 
{2\over3} \; {a(t_*)\over a(t_0)} \;  \ell_{\mathrm photon-horizon}(t_0).
\end{equation}
This formulation of the observed ``horizon constraint'' is as
model-independent as we can make it --- this constraint is a purely
kinematical statement of the observational data and is not yet a
``problem''; even in standard cosmology it will not become a problem
until one uses {\em dynamics} to deduce a specific model for
$a(t)$. In the present VSL context we will need to choose or deduce
dynamics for both $a(t)$ and $c(t)$ before this constraint can be used
to discriminate between acceptable and unacceptable cosmologies. More
on this point below.

%------------------------------------
\subsection{Monopoles and Relics}
%------------------------------------

The Kibble mechanism predicts topological defect densities that are
inversely proportional to powers of the correlation length of the
Higgs fields.  These are generally bounded above by the particle
horizon at the time of defect formation.

To simplify the analysis it is useful to use the related 
concept of Hubble distance
\begin{equation}
R_{\mathrm Hubble} = {\cp \over H}.
\end{equation}
The above quantity (often known as the Hubble radius or, speaking
loosely, `the horizon') is often mistakenly {\em identified} with the
particle horizon~\cite{Ellis-Rothman}. The two concepts, though
related, are distinct. In particular the Hubble scale evolves in the
same way as the particle horizon in simple FLRW models and hence
measures the domain of future influence of an event in these
models~\cite{Causal}.  If fields interact only through gravity, then
the Hubble scale {}\emph{is} useful as a measure of the minimum
spatial wavelength of those modes that are effectively ``frozen in''
by the expansion of the universe. A mode is said to be ``frozen in''
if its frequency is smaller than the Hubble parameter, since then
there is not enough time for it to oscillate before the universe
changes substantially, the evolution of that mode is governed by the
expansion of the universe.  Therefore, for modes travelling at the
speed $\cp$, if the ``freeze out'' occurs at $\omega<H$, this implies
that $\lambda> \cp/H$, as claimed above. Note that this discussion
crucially assumes that only gravity is operating. As soon as
interactions between fields are allowed, such as occurs in
inflationary reheating, the Hubble scale is irrelevant for determining
the evolution of modes and modes with $k/aH\ll 1$ can evolve extremely
rapidly without violating causality, as indeed typically occurs in
preheating~\cite{Causpreheat}.

If we suppose a good thermal coupling between the photons and the
Higgs field to justify using the photon horizon scale in the Kibble
freeze-out argument then we can argue as follows: Inflation
solves the relics puzzle by diluting the density of defects to an
acceptable degree, $\chi$VSL models deal with it by varying $c$ in
such a way as to make sure that the photon horizon scale is large when
the defects form.  Thus, we need the transition in the speed of light
to happen \emph{after} the spontaneous symmetry breaking (SSB) that 
leads to monopole production.

Alternatively, we could arrange a model where both photons and
the Higgs field couple directly to $\gem$, along the lines of
$S_{III}$ above; this obviates the need for postulating good thermal
coupling since the Higgs field, and its dynamics, is now directly
controlled by the variable speed of light.

So far the discussion assumes thermal equilibrium, but one should
develop a formalism which takes into account the non-equilibrium
effects and the characteristic time scales (quench and critical
slowing down scales). As a first remark one can note that the larger
the Higgs correlation length $\xi_\Phi$ is, the lower the density of
defects (with respect to the standard estimates) will be. This
correlation length characterizes the period \emph{before} the
variation of the speed of light, when we suppose that the creation of
topological defects has taken place. Remember that in the Zurek
mechanism $\rho_{\mathrm defects}\sim \xi_{\Phi}^{-n}$ with $n=1,2,$
and $3,$ for domain walls, strings, and monopoles,
respectively ~\cite{Zurek}.

We could also consider the possibility that the change in $c$ is
driven by a symmetry breaking (Higgs-like) mechanism, and try to
relate changes in $c$ to symmetry breaking at the GUT or electro-weak
scale. Unfortunately such considerations require a much more specific
model than the one considered here, and we want keep the discussion as
general as possible.

%-----------------------------------------------
\subsection{$\Lambda$ and the Planck problem}
%-----------------------------------------------

In this $\chi$VSL approach we are not affecting the cosmological
constant $\Lambda$, except indirectly via ${\cal L}_{\mathrm VSL}$.
The vacuum energy density is given by
\begin{equation}
\rho_{\Lambda}=\frac{\Lambda \; c^{2}}{8\pi\; G_{\mathrm Newton}}.
\label{rlam}
\end{equation}
But which is the $c$ appearing here? The speed of light $\cp$?  Or the
speed of gravitons $\cg$?  In our two-metric approach it is clear that
for any fundamental cosmological constant one should use $\cg$.
On the other hand, for any contribution to the total cosmological
constant from quantum zero-point fluctuations (ZPF) the situation is
more complex. If the quantum field in question couples to the metric
$\gem$, one would expect $\cp$ in the previous equation, not least in
the relationship between $\rho_{\mathrm zpf}$ and $p_{\mathrm zpf}$.

While we do nothing to mitigate the cosmological constant problem we
also do not encounter the ``Planck problem'' considered by Coule
\cite{Coule}.  He stressed the fact that in earlier VSL formulations
\cite{Moffat98,Albrecht98,Barrow98c} a varying speed of light also
affects the definition of the Planck scale. In fact, in the standard
VSL one gets two different Planck scales (determined by the values of
$c$ before and after the transition). The number of Planck times
separating the two Planck scales turns out to be larger than the
number of Planck times separating us from the standard Planck era.
So, in principle, the standard fine-tuning problems are even worse in
these models.

In contrast, in our two-metric formulation one has to decide from the
start which $c$ is referred to in the definition of the Planck length.
The definition of the Planck epoch is the scale at which the
gravitational action becomes of the order of $\hbar$. This process
involves gravity and does not refer to photons. Therefore, the $c$
appearing there is the speed of propagation of gravitons, which is
unaffected in our model.  Hence we have a VSL cosmology without a
``Planck problem'', simply because we have not made any alterations to
the gravity part of the theory.

%-----------------------------------------------------------------------
\subsection{Primordial Fluctuations}\label{partprod}
%----------------------------------------------------------------------

The inflationary scenario owes its popularity not just to its ability to
solve the main problems of the background cosmology. It is also important
because it provides a plausible, causal, micro-physics explanation for the
origin of the primordial perturbations which may have seeded large-scale
structure.  The phase of quasi-de~Sitter expansion excites the quantum
vacuum and leads to particle creation in squeezed states. As the expansion
is almost exactly exponential, these particles have an (almost exactly)
scale-invariant spectrum with amplitude given by the Hawking
``temperature" $H/2\pi$ \cite{Hu}.

In the case of {\VSL} the creation of primordial fluctuations is again
generic. The basic mechanism can be understood by modelling the change
in the speed of light as a changing ``effective refractive index of
the EM vacuum''. In an FLRW background
\begin{equation}
n_{\mathrm{em}}=\frac{\cg}{\cp}=
{1\over\sqrt{[1+(A\,M^{-4})(\partial_{t}\chi)^2]}}.
\label{neff}
\end{equation}
Particle creation from a time-varying refractive index is a well-known
effect \cite{Yab,Qed1,gamma,Prl}~\footnote{
%------------------------------------------------------------------------
It is important to stress that in the quoted papers the change of
refractive index happens in a flat static spacetime.  It is
conceivable and natural that in an FLRW spacetime the expansion rate
could play an important additional role. The results of
\cite{Yab,Qed1,gamma,Prl} should then be considered as precise in the
limit of a rapid (${\dot n}/n \gg {\dot a}/a$) transition in the speed
of light.}
%------------------------------------------------------------------------
and shares many of the features calculated for its inflationary
counterpart (\eg, the particles are also produced as squeezed
couples). We point out at this stage that these mechanisms are not
identical. In particular, in {\VSL} cosmologies it is only the fields
coupled to the EM metric that will primarily be excited. Of course, it
is conceivable, and even likely, that perturbations in these fields
will spread to the others whenever some coupling exists. Gravitational
perturbations could be efficiently excited if the $\chi$ field is
non-minimally coupled to gravity.

A second, and perhaps more fundamental, point is that a
scale-invariant spectrum of metric fluctuations on large scales is by
no means guaranteed. The spectrum may have a nearly thermal
distribution over those modes for which the adiabatic limit holds
($\tau \omega>1$, where $\tau$ is the typical time scale of the
transition in the refractive index)~\cite{gamma}.  If we assume that
$\tau$ is approximately constant in time during the phase transition,
then it is reasonable to expect an approximately Harrison--Zel'dovich
spectrum over the frequencies for which the adiabatic approximation
holds. Extremely short values of $\tau$, or very rapid changes of
$\tau$ during the transition, would be hard to make compatible with
the present observations.  Since a detailed discussion of the final
spectrum of perturbations in {\VSL} cosmologies would force us to take
into account the precise form of the $\chi$-potential $V(\chi)$,
(being very model dependent), we will not discuss these issues further
here.

As final remarks we want to mention a couple of generic features of
the creation of primordial fluctuations in {\VSL} cosmologies.  Since
we require inflation to solve the flatness problem, the {\VSL}
spectrum must be folded into the inflationary spectrum as occurs in
standard inflation with phase transitions (see \eg, \cite{Julien}).
In addition to this also a preheating phase is conceivable in {\VSL}
models if $\chi$ oscillates coherently. This would lead to production
of primordial magnetic fields due to the breaking of the conformal
invariance of the Maxwell equations.

%-----------------------------------------
\section{Flatness}
%------------------------------------------

The flatness problem is related to the fact that in FLRW cosmologies
the $\Omega=1$ solution appears as an unstable point in the evolution
of the universe. Nevertheless observations seem to be in favour of
such a value. In this section we will show that any two-metric
implementation of the kind given in equation (\ref{chimodel}) does not
by itself solve the flatness problem, let alone the quasi-flatness
problem \cite{Barrow98a}. We will also explain how this statement is
only apparently in contradiction with the claims made by Clayton and
Moffat in their implementations of two-metric VSL theories. Finally we
will show that {\VSL} can nevertheless enhance any mild SEC violation
originated by an inflaton field coupled to $\gem$.

%-----------------------------------------
\subsection{Flatness in ``pure'' {\VSL} cosmologies}
%------------------------------------------

The question ``Which $c$ are we dealing with?'' arises once more when
we address the flatness problem.  ~From the Friedmann equation we can
write
\begin{equation}
\epsilon\equiv\Omega-1=
\frac{K\;c^2}{H^{2}\;a^{2}}
=\frac{K \;c^{2}}{\dot{a}^2},
\end{equation}
where $K = 0, \pm1$. We already know that one cannot simply replace $c
\to \cp$ in the above equation. The Friedmann equation is obtained by
varying the Einstein--Hilbert action.  Therefore, the $c$ appearing
here must be the fixed $\cg$, otherwise the Bianchi identities are
violated and Einstein gravity loses its geometrical interpretation.
Thus, we have
\begin{equation}
\epsilon=\frac{K \;\cg^{2}}{\dot{a}^2}.
\end{equation} 
If we differentiate the above equation, we see that purely on
\emph{kinematic} grounds
\begin{equation}
\dot{\epsilon}
=
- 2K \;\cg^{2} \left(\frac{\ddot{a}}{\dot{a}^{3}}\right) 
= 
-2\epsilon \left(\frac{\ddot{a}}{\dot{a}}\right).
\end{equation}
{From} the way we have implemented VSL cosmology (two-metric model),
it is easy to see that this equation is independent of the photon
sector; it is unaffected if $\cp\neq\cg$.  The only way that VSL
effects could enter this discussion is indirectly. When $\cp\neq\cg$
the photon contribution to $\rho$ and $p$ is altered.

In particular, if we want to solve the flatness problem by making
$\epsilon=0$ a stable fixed point of the evolution (at least for some
portion in the history of the universe), then we must have $\ddot a >
0$, and the expansion of the universe must be accelerating (for the
same portion in the history of the universe).

It is well known that the condition $\ddot a > 0$ leads to violations
of the SEC \cite{Science}. Namely, violations of the SEC are directly
linked to solving the flatness problem.
%------------------------------------------------------
(It is for this reason that a positive cosmological constant, which
violates the SEC, is so useful in mitigating the flatness problem.)
%-----------------------------------------------------
By making use of the Friedmann equations (\ref{f1}, \ref{f2}), this
can be rephrased as
\begin{equation}
\dot{\epsilon}
=
2\epsilon 
\left[
\frac{ 
{
4\pi\;G_{\mathrm Newton}} \; \sum_{i}(\varepsilon_{i}+3p_{i}) 
}{
3H\cg^2
}
\right].
\end{equation}
In our bi-metric formalism the photon energy density $\varepsilon$ and
photon pressure $p$ are both positive, and from equation (\ref{eosp})
it is then clear that also $\varepsilon+3p$ will be positive. This is
enough to guarantee no violations of the SEC.
%---------------------------------
This means that bi-metric VSL theories are no better at solving the
flatness problem than standard cosmological (non-inflationary) FLRW
models. To ``solve'' the flatness problem by making $\epsilon=0$ a
stable fixed point will require some SEC violations and cosmological
inflation from other non-photon sectors of the theory.

%-------------------------------------------------------------------
\subsection{Flatness in the Clayton--Moffat scenarios}\label{moffat}
%-------------------------------------------------------------------

In relation to the preceding discussion, we now wish to take some time
to distinguish our approach from that of
Moffat~\cite{Moffat93a,Moffat93b,Moffat98} and
Clayton--Moffat~\cite{Clayton-Moffat-1,Clayton-Moffat-2}.  The two
clearest descriptions (of two separate VSL implementations, a
vector-based approach and a scalar-based approach) appear in the
recent papers \cite{Clayton-Moffat-1,Clayton-Moffat-2}.

%--------------------------------------------
\subsubsection{The vector scenario}
%--------------------------------------------

Let us first consider Clayton and Moffat's \emph{vector} scenario as
discussed in \cite{Clayton-Moffat-1}. In this paper Clayton and Moffat
claim to be able to solve the flatness problem directly from their VSL
implementation (equivalent to asserting that they can induce SEC
violations), an assertion we believe to be premature. The key
observation is that from their equation (6), and retaining (as much as
possible) their notation for now, it is easy to see that
\begin{eqnarray}
\rho_{\mathrm eff} &=& {\tilde\rho_\matter\over\sqrt{1+\beta \psi_0^2}}
 + {1\over2} m^2 \frac{\psi_0^2}{c^2},
\\
p_{\mathrm eff} &=& \sqrt{1+\beta \psi_0^2} \; \tilde{p}_\matter + {1\over2} m^2 
\frac{\psi_0^2}{c^2},
\\
(\rho+3p)_{\mathrm eff}  &=& 
\left(
{\tilde\rho_\matter\over\sqrt{1+\beta \psi_0^2}} +
3\;\tilde{p}_\matter\;\sqrt{1+\beta \psi_0^2}
\right) 
\nonumber\\
&& + 2 m^2 \frac{\psi_0^2}{c^2}.
\label{E:cm1}
\end{eqnarray}
[Compare also with equations (\ref{E:acceleration}) and
(\ref{E:seesaw}) below.]  Note that because the presentation in
\cite{Clayton-Moffat-1} is set up in a language where $\cp$ is kept
fixed and $\cg$ is allowed to vary, there are potential translation
pitfalls in comparing that presentation to out own approach. Here
$\tilde\rho_\matter$ and $\tilde{p}_\matter$ are the matter energy
density and pressure as measured in an orthonormal frame adapted to
the electromagnetic metric; they are simply called $\rho$ and $p$ in
the Clayton--Moffat paper.%
%--------------------------------------------------------
\footnote{We wish to thank M.A.~Clayton and J.W.~Moffat
for helpful comments on these translation issues.}
%--------------------------------------------------------

The key observation is now that contribution to the SEC arising from
the VSL vector field is positive, and if the ordinary matter has
positive pressure and energy density, then there is no possibility of
violating the SEC.  This is perhaps a little easier to see if (as is
usual in the rest of the current paper) we go to an orthonormal frame
adapted to the gravity metric, in that case
\begin{eqnarray}
\rho_{\mathrm eff} &=& \rho_\matter + 
{1\over2} m^2 \frac{\psi_0^2}{c^2},
\\
p_{\mathrm eff} &=& p_\matter + 
{1\over2} m^2 \frac{\psi_0^2}{c^2},
\\
(\rho+3p)_{\mathrm eff}  &=& \; (\rho_\matter+3p_\matter) 
+ 2 m^2 \frac{\psi_0^2}{c^2}.
\label{E:cm2}
\end{eqnarray}
The contribution to the SEC arising from the VSL vector field is
manifestly positive, and because of the form of the stress-energy
tensor, it is clear that the VSL vector field does not mimic a
cosmological constant. Again, if the ordinary matter has positive
pressure and energy density, then there is no possibility of violating
the SEC.

%--------------------------------------------
\subsubsection{The scalar scenario}
%--------------------------------------------

In Clayton and Moffat's \emph{scalar} scenario \cite{Clayton-Moffat-2}
the discussion of the relationship between SEC violations is more
nuanced, and we find ourselves largely in agreement with the point of
view presented in that paper.  Indeed, subtract equation (30) of that
paper from equation (31) and divide by two to obtain (following the
notation of that paper)
\begin{eqnarray}
\label{E:acceleration}
{\ddot R\over R} &=& {1\over 3} c^2 \Lambda + {1\over3} c^2 V(\phi) 
-{1\over 6} \dot \phi^2 - {\kappa c^2\over 6} 
\left( {\rho_M\over\sqrt{I}} + 3 p_M \sqrt{I} \right).
\nonumber\\
&&
\end{eqnarray}
The quantity $I$ is defined in equation (15) of that paper and
satisfies $I>1$, so that the square root is well defined, ($\sqrt{I}
\mapsto \cp/\cg$ when mapped to our notation.) If the ``ordinary''
matter ($\rho_M$, $p_M$) is indeed ``ordinary'' ($\rho_M>0$,
$p_M>0$), the only possible source of SEC violations (and
inflation) is from the explicit cosmological constant or from letting
the VSL field ($\phi$ in their notation, which becomes $\chi$ in ours)
act as an inflaton field. Alternatively, if $p_M$ is slightly negative
and $I$ is large, the effect of this negative pressure is greatly
enhanced, possibly leading to SEC violations.

We conclude from the previous discussion that two-metric VSL
cosmologies do not automatically solve the flatness problem --- to
solve the flatness problem one needs to make the universe expand
rapidly, which means that there are SEC violations (with respect to
the \emph{gravity} metric).

Though we disagree with Clayton and Moffat on the technical issue of
whether two-metric VSL cosmologies can automatically solve the
flatness problem, we do wish to emphasise that we are largely in
agreement with those papers on other issues --- in particular, we
strongly support the two-metric approach to VSL cosmologies.
Furthermore, as we will now discuss, we agree that two-metric VSL
cosmologies naturally lead to an amplification of any inflationary
tendencies that might be present in those fields that couple to the
photon metric.

%-------------------------------------------------------------------
\subsection{Flatness in Heterotic (Inflaton+\VSL) models.}
\label{hete}
%-------------------------------------------------------------------

To conclude this section we will show how two-metric VSL cosmologies
\emph{enhance} any inflationary tendencies in the matter sector. Let
us suppose that we have an inflaton field coupled to the
\emph{electromagnetic} metric. We know that during the inflationary
phase we can approximately write
\begin{equation}
T^{\mu\nu}_{\mathrm{inflaton}} \propto g_{\mathrm em}^{\mu \nu}.
\end{equation}
We have repeatedly emphasized that it is important to define
\emph{the} physical energy density and pressure ($\varepsilon, p$) as
the appropriate components of the stress-energy tensor when referred
to an orthonormal basis \emph{of the metric that enters the Einstein
equation}.  The condition $T^{\mu\nu}_{\mathrm inflaton} \propto
g^{\mu\nu}_{\mathrm em}$, when expressed in terms of an orthonormal
basis of the metric $g$ asserts
\begin{equation}
 p_{\mathrm inflaton} = - {\cp^2\over\cg^2} \; \varepsilon_{\mathrm inflaton}.
\end{equation}
That is
\begin{equation}
(\varepsilon + 3p)_{\mathrm inflaton} = 
\left(1-3 {\cp^2\over\cg^2}\right) \varepsilon_{\mathrm inflaton}.
\end{equation}
Thus, any ``normal'' inflation will be amplified during a VSL
epoch. It is in this sense that VSL cosmologies heterotically improve
standard inflationary models.

We can generalize this argument. Suppose the ``normal'' matter, when
viewed from an orthonormal frame adapted to the \emph{electromagnetic}
metric, has energy density $\tilde\varepsilon$ and pressure $\tilde
p$.  {From} our previous discussion [equations
(\ref{E:two-stress-tensors-FLRW})---(\ref{E:p})] we deduce
\begin{eqnarray}
\varepsilon+3p &=& 
{\cg\over\cp} \; \tilde\varepsilon + 3 \; {\cp\over\cg} \; \tilde p.
\label{E:seesaw}
\end{eqnarray}
[Compare with equations (\ref{E:cm1}) and (\ref{E:acceleration})
above.]  In particular, if $\tilde p$ is slightly negative, VSL
effects can magnify this to the point of violating the SEC (defined
with respect to the gravity metric). It is in this sense that
two-metric VSL cosmologies provide a natural enhancing effect for
negative pressures (possibly leading to SEC violations), even if they
do not provide the seed for a negative pressure.

We point out that this same effect makes it easy to violate \emph{all}
the energy conditions. If ($\tilde\varepsilon,\tilde p$) satisfy all
the energy conditions with respect to the photon metric, and provided
$\tilde p$ is only slightly negative, then VSL effects make it easy
for ($\varepsilon,p$) to violate all the energy conditions with
respect to the gravity metric --- and it is the energy conditions with
respect to the gravity metric that are relevant to the singularity
theorems, positive mass theorem, and topological censorship theorem.

%---------------------------------------------------------------------------
\section{The entropy problem}
%---------------------------------------------------------------------------

It is interesting to note that (at least in the usual framework)
the two major cosmological puzzles described above (isotropy/horizon
and flatness) can be reduced to a single problem related to the huge
total amount of entropy that our universe appears to have
today~\cite{Guth,300,Hu0,LF}. If we define $s\propto T^3$ the entropy
density associated with relativistic particles and $S=a^{3}(t)s$ the
total entropy per comoving volume, then it is easy to see from the
Friedmann equation (\ref{f1}) that
\begin{equation}
  \label{eq:flrwent}
   a^2=\frac{K \cg^2}{H^2 (\Omega-1)},
\end{equation}
and so
\begin{equation}
  \label{eq:sflrw}
  S=\left[ \frac{K \cg^2}{H^2\left(\Omega-1\right)}\right]^{3/2} s.
\end{equation}
The value of the total entropy can be evaluated at the present time
and comes out to be $S>10^{87}$. One can then see that explaining why
$\Omega\approx 1$ (the flatness problem) is equivalent to explaining
why the entropy of our universe is so huge.

In a similar way one can argue (at least in the usual framework) that
the horizon problem can be related to the entropy
problem~\cite{Guth,300,Hu0}. In order to see how large the causally
connected region of the universe was at the time of decoupling with
respect to our present horizon, we can compare the particle horizon at
time $t$ for a signal emitted at $t=0$, $\ell_{\mathrm h}(t)$, with
the radius at same time, $L(t)$, of the region which now corresponds
to our observed universe of radius $L_{\mathrm{present}}$. The fact
that (assuming insignificant entropy production between decoupling and
the present epoch) $(\ell_{\mathrm h}/L)^{3}
|_{t_{\mathrm{decoupling}}} \ll 1$ is argued to be equivalent to the
horizon problem.  Once again, a mechanism able to greatly increase $S$
via a non-adiabatic evolution would also automatically lead to the
resolution of the puzzle.

{\VSL} cosmologies evade this connection between the horizon and
flatness puzzles: We have just seen that although the horizon problem
is straightforwardly solved, it is impossible to solve  the
flatness dilemma (at least in pure {\VSL} models). To understand how
this may happen is indeed very instructive.

First of all, we can try to understand what happens to the entropy per
comoving volume $S=a^{3}(t)s$. In the case of inflation we saw that
the non-adiabatic evolution $\dot{S}\neq 0$ was due to the fact that
although the entropy densities do not significantly change,
$s_{\mathrm{before}} \approx s_{\mathrm{after}}$ thanks to reheating,
nevertheless the enormous change in scale factor
$a(t_{\mathrm{after}}) = \exp[
H(t_{\mathrm{after}}-t_{\mathrm{before}}) ] \cdot
a(t_{\mathrm{before}})$ drives an enormous increase in total entropy
per comoving volume.  (Here ``before'' and ``after'' are intended with
respect to the inflationary phase.)

In our case (bimetric VSL models) the scale factor is unaffected by
the transition in the speed of light  if the $\chi$ field is not the 
dominant energy component of the universe. Instead what changes is the
entropy density $s$.  As we have seen, a sudden phase transition
affecting the speed of light induces particle creation and raises both
the number and the average temperature of relativistic particles.
Therefore one should expect that $s$ grows as $\cp\to\cg$.

From equation (\ref{E:horizon}) it is clear that the increased speed
of light is enough to ensure a resolution of the horizon problem,
regardless of what happens to the entropy. At the same time one can
instead see that the flatness problem is not solved at all. Equation
(\ref{eq:sflrw}) tells us that it is the {\em ratio} $S\;H^3/s\approx
\dot{a}^3$ which determines the possibility of stretching the
universe. Unfortunately this is not a growing quantity in the standard
model as well as in {\em pure} bi-metric VSL theory. Once again only
violations of the SEC ($\ddot{a}>0$) can lead to a resolution of the
flatness problem.

%------------------------------------------------------------------------
\section{Observational tests and the low-redshift {\VSL} universe}
\label{obst}
%------------------------------------------------------------------------

At this point, it is important to note that due to the nature of the
interaction (\ref{s1}), the $\chi$ field appears unable to decay
completely. Decay of the $\chi$ field proceeds via $2\chi \rightarrow
2\gamma$ and hence, once the density of $\chi$ bosons drops
considerably, ``freeze-out'' will occur and the $\chi$ field will stop
decaying.  This implies that the $\chi$ field \emph{may} be
dynamically important at low-redshift \emph{if} its potential is such
that its energy density drops less rapidly than that of radiation.

However, the $\chi$ correction to $g_{em}$ corresponds to a dimension
twelve operator, which is highly non-renormalizable.  The vector model
of Moffat \cite{Clayton-Moffat-1} is a dimension eight operator.
Nevertheless, for energies below $M$ it is difficult to argue why
either of these operators will not be negligibly small relative to
dimension five operators, which would cause single body decays of the
$\chi$ field. While it is possible that these dimension five operators
are absent through a global symmetry \cite{Carroll-1}, or the lifetime
of the $\chi$ bosons is extremely long, we will see later that such
non-renormalizable interactions with the standard model give rise to
serious constraints. For the time being we neglect single-body decays,
and we can imagine two natural dark-matter candidates, with the added
advantage that they are distinguishable and detectable, at least in
principle.

(i) If $V(\chi)$ has a quadratic minimum, the $\chi$ field will
oscillate about this minimum and its average equation of state will be
that of dust. This implies that the $\chi$ field will behave like
axions or cold dark matter. Similarly if the potential is quartic, the
average equation of state will be that of radiation.

(ii) If $V(\chi)$ has quintessence form, with no local minimum but a
global minimum at $\chi \rightarrow \infty$. A typical candidate is a
potential which decays to zero at large $\chi$ (less rapidly than an
exponential) with $V(\chi) > A e^{-\lambda \chi}$ for $\lambda > 0$.

These two potentials lead to interesting observational implications
for the low-redshift universe which we now proceed to analyze and
constrain.

%------------------------------------------------------------------------
\subsection{Clustering and gravitational lensing}
%------------------------------------------------------------------------

It is interesting to note that the effective refractive index we
introduced in equation (\ref{neff}) may depend, not just on time, but
also on space and have an anisotropic structure.  In particular the
dispersion relation of photons in an anisotropic medium reads
\begin{equation}
\omega^{2}= [n^{-2}]^{ij} \; k_{i} \; k_{j},
\end{equation}
and from the above expression it is easy to see that the
generalization of equation (\ref{neff}) then takes the form
\begin{equation}
[n^{-2}]^{ij}=g^{ij}_{\mathrm em}/|g^{tt}_{\mathrm em}|.
\label{E:aneff}
\end{equation}

Scalar fields do not support small scale density inhomogeneities
(largely irrespective of the potential). This implies that the
transfer function tends to unity on small scales and the scalar field
is locally identical to a cosmological constant.

However, on scales larger than $100 Mpc$, the scalar field can cluster
\cite{Ma}. During such evolution both $\dot\chi \neq 0$ and
$\partial_i \chi \neq 0$ will hold. This would lead to deviations from
equation (\ref{neff}), as the ratio between the two speeds of light
will not be only a function of time.

{For} instance, let us suppose we are in a regime where time
derivatives of $\chi$ can be neglected with respect to spatial
derivatives. Under these conditions the EM metric reduces to
\begin{eqnarray}
g^{\mathrm em}_{tt} &=& g_{tt} = - |g_{tt}|, 
\\
g^{\mathrm em}_{ij} &=&
g_{ij}-(AM^{-4}) \; \partial_{i}\chi \; \partial_{j}\chi.
\end{eqnarray}  

{From} equation (\ref{E:aneff}) this is equivalent to a tensor refractive
index $n_{ij}$, with
\begin{equation}
[n^{2}]_{ij}=
{g_{ij}-(AM^{-4}) \; \partial_{i}\chi \; \partial_{j}\chi 
\over 
|g_{tt}|}. 
\end{equation}
This tensor refractive index may lead to additional lensing by
large-scale structure, over and above the usual contribution from
gravitational lensing \cite{SEF}.

%------------------------------------------------------------------------
\subsection{Quintessence and long-range forces}
%------------------------------------------------------------------------

Another natural application is to attempt to use the $\chi$ field as
the source of the ``dark energy'' of the universe, the putative source
of cosmic acceleration. This is attractive for its potential to unify
a large number of disparate ideas, but is severely constrained as
well.

%------------------------------------------------------------------------
\subsubsection{Constraints arising from variation of the fine-structure
constant} 
%------------------------------------------------------------------------

As noted in the introduction, a change of $c_{\mathrm photon}$ will
cause a variation in the fine-structure constant. Such variation is
very constrained. We point out two particularly interesting
constraints.  The first, arising from nucleosynthesis \cite{KPW}, is
powerful due to the extreme sensitivity of nucleosynthesis to
variations in the proton-neutron mass difference, which in turn is
sensitive to $\alpha$. This places the tight constraint that $
|\dot{\alpha}/\alpha| \leq 10^{-14} yr^{-1}$. However, this is only a
constraint on $\dot{c}_{\mathrm photon}/c_{\mathrm photon}$ if no
other constants appearing in $\alpha$ are allowed to vary. Further we
have assumed $\dot{\alpha}$ was constant through nucleosynthesis.

A similar caveat applies to other constraints one derives for
variations of $c_{\mathrm photon}$ through variations of
$\alpha$. Other tests are only sensitive to integrated changes in
$\alpha$ over long time scales. At redshifts $z \leq 1$ constraints
exist that $|\Delta{\alpha}/\alpha| < 3 \times 10^{-6}$ (quasar
absorption spectra \cite{QSO}) and $|\Delta{\alpha}/\alpha| < 10^{-7}$
(Oklo natural reactor \cite{oklo}).

%------------------------------------------------------------------------
\subsubsection{Binary pulsar constraints} 
%------------------------------------------------------------------------

Unless we choose the unattractive solution that $\chi$ lies at the
minimum of its potential but has non-zero energy (\ie, an explicit
$\Lambda$ term), we are forced to suggest that $\dot{\chi} \neq 0$
today and $V(\chi)$ is of the form $e^{-\lambda \chi}$ or $\chi^{-n}$
\cite{Quin}. In this case, gravitons and photons do not travel at the
same speed today. The difference in the two velocities is rather
constrained by binary pulsar data to be less than $1\%$ \cite{Carlip};
\ie, $|n_{\mathrm{em}} - 1| < 0.01$.

%------------------------------------------------------------------------
\subsubsection{High-energy tests of VSL}
%------------------------------------------------------------------------

Constraints on our various actions $S_I - S_{III}$ also come from high
energy experiments. In model $I$, photons travel faster than any other
fields. This would lead to perturbations in the spectrum of nuclear
energy levels \cite{D96}.

Similarly, high energy phenomena will be sensitive to such speed
differences. For example, if $c_{\mathrm photon} > c_{e^-}$, the
process $\gamma \rightarrow e^- + e^+$ becomes kinematically possible
for sufficiently energetic photons. The observation of primary cosmic
ray photons with energies up to 20 TeV implies that today $c_{\mathrm
photon} - c_{e^-} < 10^{-15}$ \cite{CG}.  The reverse possibility ---
which is impossible in our model I if $A > 0$ in equation
(\ref{cphoton}) --- is less constrained, but the absence of vacuum
Cherenkov radiation with electrons up to 500 GeV implies that $c_{e^-}
- c_{\mathrm photon} < 5 \times 10^{-13}$. Similar constraints exist
which place upper limits on the differences in speeds between other
charged leptons and hadrons \cite{CG,cptlorentz}. These will generally
allow one to constrain models I -- III, but we will not consider such
constraints further.
 
%------------------------------------------------------------------------
\subsubsection{Non-renormalizable interactions with the standard model}
%------------------------------------------------------------------------

Our $\chi$VSL model is non-renormalizable and hence one expects an
infinite number of $M$-scale suppressed, dimension five and higher,
interactions of the form
\beq
\beta_i \frac{\chi^n}{M^{n}}{\cal L}_i,
\label{nonrenc}
\eeq
where $\beta_i$ are dimensionless couplings of order unity and ${\cal
L}_i$ is any dimension-four operator such as $F^{\mu \nu} F_{\mu
\nu}$.

For sub-Planckian $\chi$-field values, the tightest constraints
typically come from $n =1$ (dimension five operators) and we focus on
this case. The non-renormalizable couplings will cause time variation
of fundamental constants and rotation of the plane of polarization of
distant sources \cite{Carroll}.  For example, with ${\cal L}_{QCD} =
\Tr(G_{\mu\nu} G^{\mu \nu})$, where $G_{\mu \nu}$ is the QCD field
strength, one finds the strict limit \cite{QCD}
\beq
|\beta_{G^2}| \leq 10^{-4} (M/M_{\mathrm{Planck}})
\eeq
which, importantly, is $\chi$ independent. 

If one expects that $|\beta_i| = O(1)$ on general grounds, then this
already provides as strong a constraint on our model as it does on
general quintessence models.  This constraint is not a problem if
there exist exact or approximate global symmetries
\cite{Carroll-1}. Nevertheless, without good reason for adopting such
symmetries this option seems unappealing.

Another dimension five coupling is given by equation (\ref{nonrenc})
with $ {\cal L}_{F^2} = F_{\mu \nu} F^{\mu \nu}$ which causes
time-variation in $\alpha$. Although there is some evidence for this
\cite{varyalpha}, other tests have been negative as discussed
earlier. These yield the constraint \cite{Carroll-1}
\beq
|\beta_{F^2}| \leq 10^{-6} (M H/\langle\dot{\chi}\rangle).
\eeq
Clearly this does not provide a constraint on {\VSL} unless we
envisage that $\dot{\chi} \neq 0$ today as required for
quintessence. If $\chi$ has been at the minimum of its effective
potential since around $z < 5$, then neither this, nor the binary
pulsar, constrain {\VSL} models.  The CMB provides a more powerful
probe of variation of fundamental constants and hence provides a test
of {\VSL} if $\chi$ did not reach its minimum before $z \simeq 1100$
\cite{Barrow-cmb}.

Another interesting coupling is ${\cal L}_{F^*F} = F_{\mu \nu}
{}^*F^{\mu\nu}$, where ${}^*F$ is the dual of $F$. As has been noted
\cite{Carroll-1}, this term is not suppressed by the exact global
symmetry $\chi\rightarrow\chi + {\mathrm constant}$, since it is
proportional to $(\nabla_{\mu}\chi) \; A_{\nu} \; {}^*F^{\mu \nu}$.  A
non-zero $\dot{\chi}$ leads to a polarization-dependent ($\pm$)
deformation of the dispersion relation for light
\begin{equation}
\omega^2 = k^2 \pm\beta_{F^*F} (\dot{\chi} k/M)\,.
\end{equation}
If $\dot{\chi} \neq 0$ today, the resulting rotation of the plane of
polarization of light traveling over cosmological distances is
potentially observable.  Indeed claims of such detection exist
\cite{Ralston}. However, more recent data is consistent with no
rotation \cite{Leahy,CF}.  Ruling out of this effect by
high-resolution observations of large numbers of sources would be
rather damning for quintessence but would simply restrict the $\chi$
field to lie at its minimum, \ie, $\Delta \chi \simeq 0$ for $z < 2$.

On the other hand, a similar and very interesting effect arises not
from $\dot{\chi}$ but from spatial gradients of $\chi$ at
low-redshifts due to the tensor effective refractive index of
spacetime.

%----------------------------------------------------------------------
\section{Discussion}
%---------------------------------------------------------------------
In this paper we have tried to set out a geometrically consistent and
physically coherent formalism for discussing Variable Speed of Light
(VSL) cosmologies. An important observation is that taking the usual
theory and simply replacing $c \to c(t)$ is {\em more radical a step
than strictly necessary}. One either ends up with a coordinate change
which does not affect the physics, or one is forced to move well
outside the usual mathematical framework of Lorentzian differential
geometry.  In particular, replacing $c \to c(t)$ in the Einstein
tensor of an FLRW universe violates the Bianchi identities and energy
conservation and destroys the usual geometrical interpretation of
Einstein gravity as arising from spacetime curvature. We do not claim
that such a procedure is necessarily wrong, but point out that it is a
serious and fundamental modification of our usual ideas.

In contrast, in the class of {\VSL} cosmologies presented in this
article, where the Lorentz symmetry is ``softly broken'', the
``geometrical interpretation'' is preserved, and the Bianchi
identities are fulfilled. In particular, these ``soft breaking'' VSL
scenarios are based on straightforward extensions of known physics,
such as the Scharnhorst effect and anomalous electromagnetic
propagation in gravitational fields, and so represent ``minimalist''
implementations of VSL theories.  Indeed, these non-renormalizable
VSL-inducing couplings should exist in supergravity theories, though
they would be expected to be negligible at low energies.

In this article, we have argued for the usefulness of a
two-metric approach. We have sketched a number of two-metric scenarios
that are compatible with laboratory particle physics, and have
indicated how they relate to the cosmological puzzles. We emphasise
that there is considerable freedom in these models, and that a
detailed confrontation with experimental data will require the
development of an equally detailed VSL model.  In this regard VSL
cosmologies are no different from inflationary cosmologies. Since the
models we discuss are non-renormalizable however, there may be
interesting implications for the low-redshift universe through
gravitational lensing and birefringence.

VSL cosmologies should be seen as a general scheme for attacking
cosmological problems. This scheme has some points in common with
inflationary scenarios, but also has some very strange peculiarities
of its own. In particular, once $\cp\neq\cg$ complications may appear
in rather unexpected places.

%--------------------------------------------------------------------
\section*{Acknowledgments}
%--------------------------------------------------------------------

BB thanks John Barrow, Kristin Burgess, Alan Guth, Roy Maartens,
Nazeem Mustapha, and Joao Magueijo for discussions and/or comments on
drafts, and the Newton Institute for hospitality.  SL wishes to thank
Carlo Baccigalupi, Julien Lesgourgues, and Sebastiano Sonego for
useful discussions and remarks.
 
We also wish to thank M.A.~Clayton and J.W.~Moffat for their interest,
and for useful discussions on translation issues.

This research was supported in part by the Newton Institute,
Cambridge, during the program ``Structure Formation in the Universe''
(BB), by the Italian Ministry of Science (SL), and by the US
Department of Energy (CMP and MV).

In addition BB, CMP, and MV wish to thank SISSA (Trieste, Italy) for
support and hospitality. CMP and MV also thank LAEFF (Laboratorio de
{\Astrofisica} Espacial y {\Fisica} Fundamental, Madrid, Spain) for
their hospitality during various stages of this research.

%------------------------------------------------------------------
\appendix
%-------------------------------------------------------------------

\newpage 
%-------------------------------------------------------------------
\section{Notation}
%-------------------------------------------------------------------

%\newpage 

\begin{table}[t]
\centerline{Summary of notation used in this article.}
%\medskip
\begin{tabular}[t]{||l|l|l||}
Symbol & Brief Description & Eqn. \\
\hline
\hline
& & \\
$\; g_{\mathrm{em}}$  & The electromagnetic metric &  \ref{chimodel} \\
$\; \epsilon$  & Energy density &  \ref{energyden}\\
$\; \rho$  & Mass density &  \ref{rho} \\
$\; p$  & Pressure &  \ref{energyden} \\
$\; c_{\mathrm{gravity}}$  & Velocity of gravitons &  \ref{zeta}  \\
$\; c_{\mathrm{photon}}$  & Velocity of photons &  \ref{cphoton}  \\
$\; c_{e^-}$  & Maximum velocity of electrons &    \\
%$\; \alpha$  & The fine-structure constant &  \ref{fs} \\
$\; \beta_{1,2,3}$ & Coefficients of 1-loop QED corrections & \ref{dh} \\
%$\; \epsilon_0$  & The permittivity of space &  \ref{fs}  \\
%$\; e$  & The electron charge &  \ref{fs}  \\
$\; \zeta$  & The ratio of photon to graviton velocity. & \ref{zeta} 
\\
$\;
 m_e$  & The electron mass  &  \ref{dh}
\\
$\;
\chi$  & The VSL-inducing field &  \ref{chimodel}
\\
$\;
\psi$  & A generic spinor field &  \ref{s2} 
\\
$\;
M$  & The scale for $\chi$ &  \ref{chimodel}
 \\
&  non-renormalization effects &  \\
$\;
A$  & The coupling constant for & \ref{chimodel}  
\\
& the interaction between $\chi$ and $F_{\mu\nu}$ & \\
$\;
K$  & The tri-curvature constant: $K = 0,\pm 1$  &  \ref{einstein}
\\
$\;
G_{\mathrm{Newton}}$  & Newton's gravitational constant  & \ref{gmod} 
\\
$\;
\rho_{\Lambda}$& The energy density in $\Lambda$ & \ref{rlam} 
\\   
$\;
\Lambda$ & The cosmological constant & \ref{rlam}
\\   
$\;
n_{\mathrm{em}}$& The effective refractive index & \ref{neff} \\ 
& of spacetime & \\   
$\;
\gamma$& A generic photon &   \\   
$\;
V(\chi)$ & The $\chi$ potential & \ref{s1} 
\\   
$\;
\omega$ & Photon frequency &  
\\   
$\; \tau$ & Timescale for the $\chi$-field phase transition & \\ 
& & \\   
\end{tabular}
%\medskip
\caption{Symbols used in the paper with a brief description and an
equation where it is first used, if applicable.}
\end{table}

%-------------------------------------------------------------------
\section{Varying $\cg$, \\ keeping $\cp$ fixed.}\label{cgvary}
%-------------------------------------------------------------------

In contrast with the main thrust of this paper, we will now ask what
happens if we keep $\cp$ fixed, while letting $\cg$ vary. This means
that we are still dealing with a two-metric theory, and so it still
makes sense to define VSL in terms of the ratio $\cp/\cg$. Keeping
$\cp$ fixed has the advantage that the photon sector (or more
generally the entire matter sector) has the usual behaviour.  However
a variable $\cg$ has the potential for making life in the gravity
sector rather difficult.

To make this model concrete, consider a relationship between the
photon metric and the gravity metric of the form
\begin{equation}
[\ggrav]_{\alpha\beta} = 
[\gem]_{\alpha\beta} +
(A\; M^{-4}) \;
\nabla_\alpha \chi \; \nabla_\beta \chi,
\end{equation}
where we now take $\gem$ as fundamental, and $\ggrav$ as the derived
quantity. We postulate an action of the form
\begin{eqnarray}
S_{IV} &=& 
\int d^4 x \sqrt{-\ggrav} \; R(\ggrav)
\nonumber\\
&+&
\int d^4 x \sqrt{-\gem} \; 
\L_{\mathrm matter}(\gem,\psi,\chi), 
\end{eqnarray}
where the matter Lagrangian now includes \emph{everything}
non-gravitational and the $\chi$ field. The matter equations of motion
are the usual ones and it makes most sense to define the stress-energy
tensor with respect to the photon metric. (That is, use $\tilde
T^{\mu\nu}$ as the primary quantity.)  The Einstein equation is
modified to read
\begin{equation}
{\sqrt{{\ggrav}\over{\gem}}} \; 
G^{\mu\nu}|_{\ggrav = \gem + A M^{-4} \nabla \chi \otimes \nabla \chi} = 
\tilde T^{\mu\nu}.
\end{equation}
Though minor technical details differ from the approach adopted in
this paper ($\cg$ fixed, $\cp$ variable), the results are
qualitatively similar to our present approach. We will for the time
being defer further discussion of this possibility.

\clearpage
%--------------------------------------------------------------------

%--------------------------------------------------------------------
\end{document}